\documentclass[onecollarge,natbib]{svjour2}
\bibpunct{[}{]}{;}{n}{}{,}
\smartqed  
\usepackage{graphicx}
\usepackage{mathptmx}
\usepackage{latexsym}
\usepackage{amsmath}
\usepackage{amssymb}
\usepackage{url}

\begin{document}
\title{Delineating the polarized and unpolarized partonic structure of the nucleon}
\titlerunning{Delineating the polarized and unpolarized partonic structure of the nucleon} 
\author{Pedro Jimenez-Delgado}
\institute{Pedro Jimenez-Delgado \at
           Thomas Jefferson National Accelerator Facility \\
           12000 Jefferson Avenue, Suite 1, Newport News, VA 23606, USA\\
           Tel.: +1-757-269-7870\\
           Fax: + 1-757-269-7002\\
           \email{pedro@jlab.org}}
\date{Received: date / Accepted: date} 
\maketitle
\begin{abstract}
Reports on our latest extractions of parton distribution functions of the nucleon are given. First an overview of the recent JR14 upgrade of our unpolarized PDFs, including NNLO determinations of the strong coupling constant and a discussion of the role of the input scale in parton distribution analysis. In the second part of the talk recent results on the determination of spin-dependent PDFs from the JAM collaboration are reported, including a careful treatment of hadronic and nuclear corrections, as well as reports on the impact of present and future data in our understanding of the spin of the nucleon. 
\end{abstract}

\section{Introduction}\label{introduction}
Perturbative quantum chromodynamics has been shown to provide an excellent description of hard scattering processes at particle accelerators. The quantitative description of high-energy collisions involving hadrons in the initial state relies on the fact that in such interactions the hadronic structure, in terms of their constituent quarks and gluons (partons), may be embodied by universal parton distribution functions (PDFs). In a first approximation the PDFs can be taken (at a resolution scale) to depend only on the momentum (fraction) of the parton parallel to that of the hadron to which it belongs (parent). Thus additional degrees of freedom, namely the transverse components of the parton momentum as well as spatial dependences, are disregarded (integrated out). This defines the so-called collinear approximation, which has been developed over the last few decades by the world community and provides the base for many analyses at current facilities.

The PDFs are typically determined by simultaneously fitting a wide variety of data for large momentum transfer processes (global analysis). The parameters of the fits describe the PDFs at some initial (input) scale, while evolution equations are then used to calculate the PDFs at all other scales needed for the calculations. Although in principle the fundamental distributions in nature are the PDFs for a specific helicity ($f_i^\uparrow$ and $f_i^\downarrow$, i.e. corresponding respectively to parton spins aligned and anti-aligned with that of the hadron), experiments with unpolarized beams and targets are sensitive only to the averaged helicity distributions or \emph{unpolarized} PDFs ($f_i = f_i^\uparrow + f_i^\downarrow$), while information on the \emph{polarized} distributions ($\Delta f_i = f_i^\uparrow - f_i^\downarrow$) can be obtained from measurements involving polarized beams and/or targets. Thus traditionally the unpolarized and polarized cases have been treated separately, although in principle one could perform a global fit of polarized and unpolarized data simultaneously. 

A comprehensive review on the determination of polarized and unpolarized PDFs has been recently presented in \cite{Jimenez-Delgado:2013sma}, in this talk we will briefly discuss some aspects of our latest extractions. We will start with the unpolarized case in Sect.~\ref{unpolarized} by discussing some aspects of the recent JR14 analysis \cite{Jimenez-Delgado:2014twa}. Section \ref{polarized} will be devoted to the polarized case and will report on the first extraction by the Jefferson Lab Angular Momentum Collaboration, JAM13 \cite{Jimenez-Delgado:2013boa}, as well as on later results on constraints at large momentum fractions by future data from Jefferson Lab at 12 GeV \cite{Jimenez-Delgado:2014xza}.

\section{Unpolarized case: JR14}\label{unpolarized}
The recent JR14 update of our distributions \cite{Jimenez-Delgado:2014twa} utilizes the world data on DIS measurements and data on Drell-Yan dilepton and (up to NLO) high-$p_T$ inclusive jet production for extracting the unpolarized parton distributions of the nucleon together with the (highly correlated) strong coupling $\alpha_s$ at NLO and NNLO of perturbative QCD. 

To ensure the reliability of our results we have included non-perturbative higher-twist (HT) terms, and nuclear corrections for the deuteron structure functions together with off-shell and nuclear shadowing corrections, as well as target-mass corrections for $F_2$ and $F_L$. To learn about the stability of the results, we performed fits to subsets of data by applying various kinematic ($Q^2, W^2$) cuts on the available data, since in particular the HT contributions turned out to be sensitive to such choices. A safe and stable choice turned out to be $Q^2\geq 2$ GeV$^2$ and $W^2\geq 3.5$ GeV$^2$, which are our final nominal cuts imposed on DIS data. Besides these improvements in the theoretical computations a novelty with respect to our previous analyses is a complete treatment of the systematic uncertainties of the data including experimental correlations. 

Yet another improvement is the use of the running-mass definition for DIS charm and bottom production, which results in an improved stability of the perturbative series. Since heavy quark coefficient functions are exactly known only up to NLO, we have refrained from using DIS heavy-quark production data and inclusive jet production data for our nominal NNLO fits; nevertheless the NLO results are in good agreement with experiment \cite{Jimenez-Delgado:2014twa}. The same holds true when fitting the data at NNLO using (theoretically inconsistent) NLO matrix elements (referred to as NNLO*) or approximate NNLO ones. It should, however, be kept in mind that the correct order of (massive) matrix elements appears to be far more important than the chosen order of PDFs.

\begin{figure*}[b]
\centering
\includegraphics[width=0.49\textwidth]{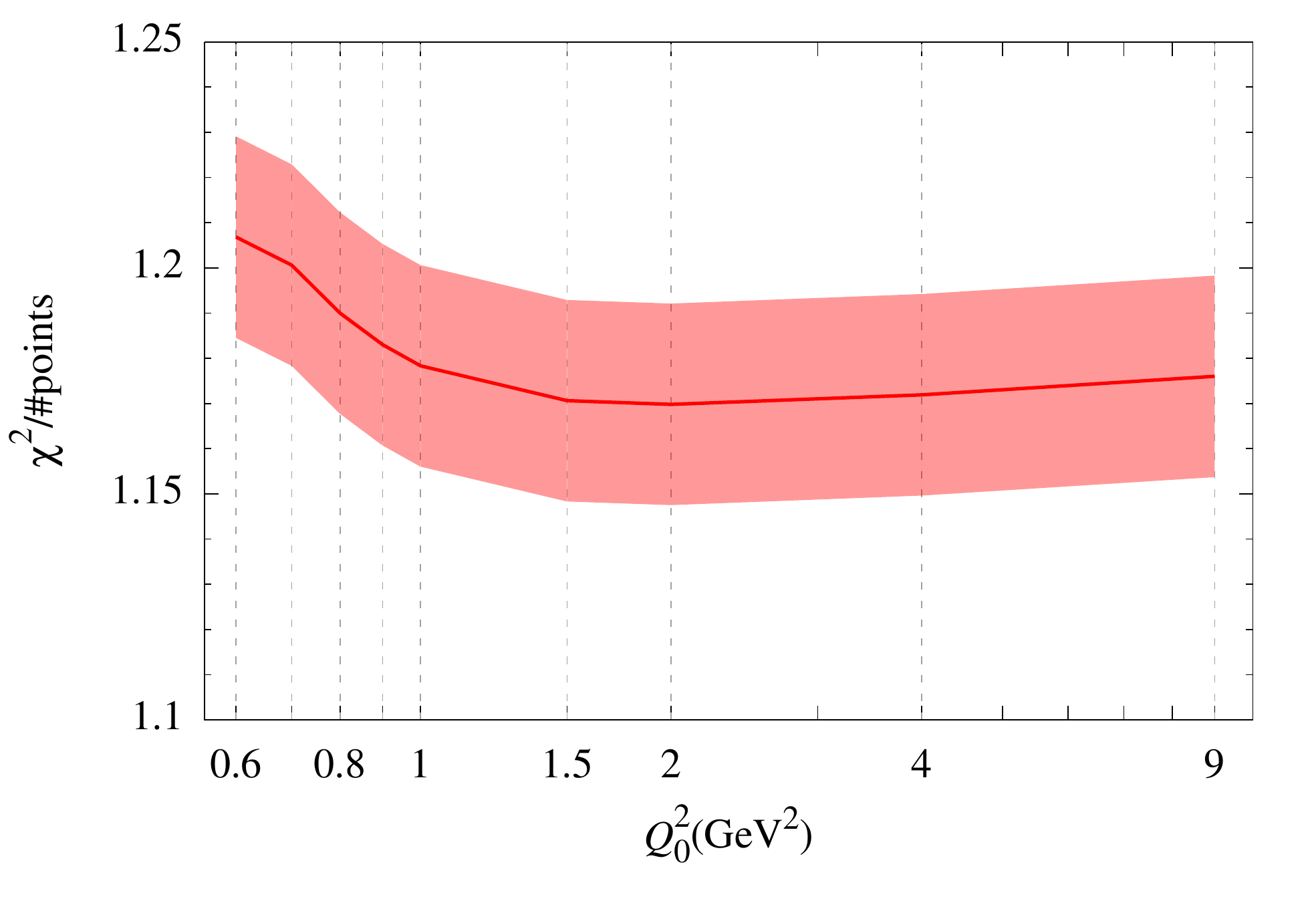}
\includegraphics[width=0.49\textwidth]{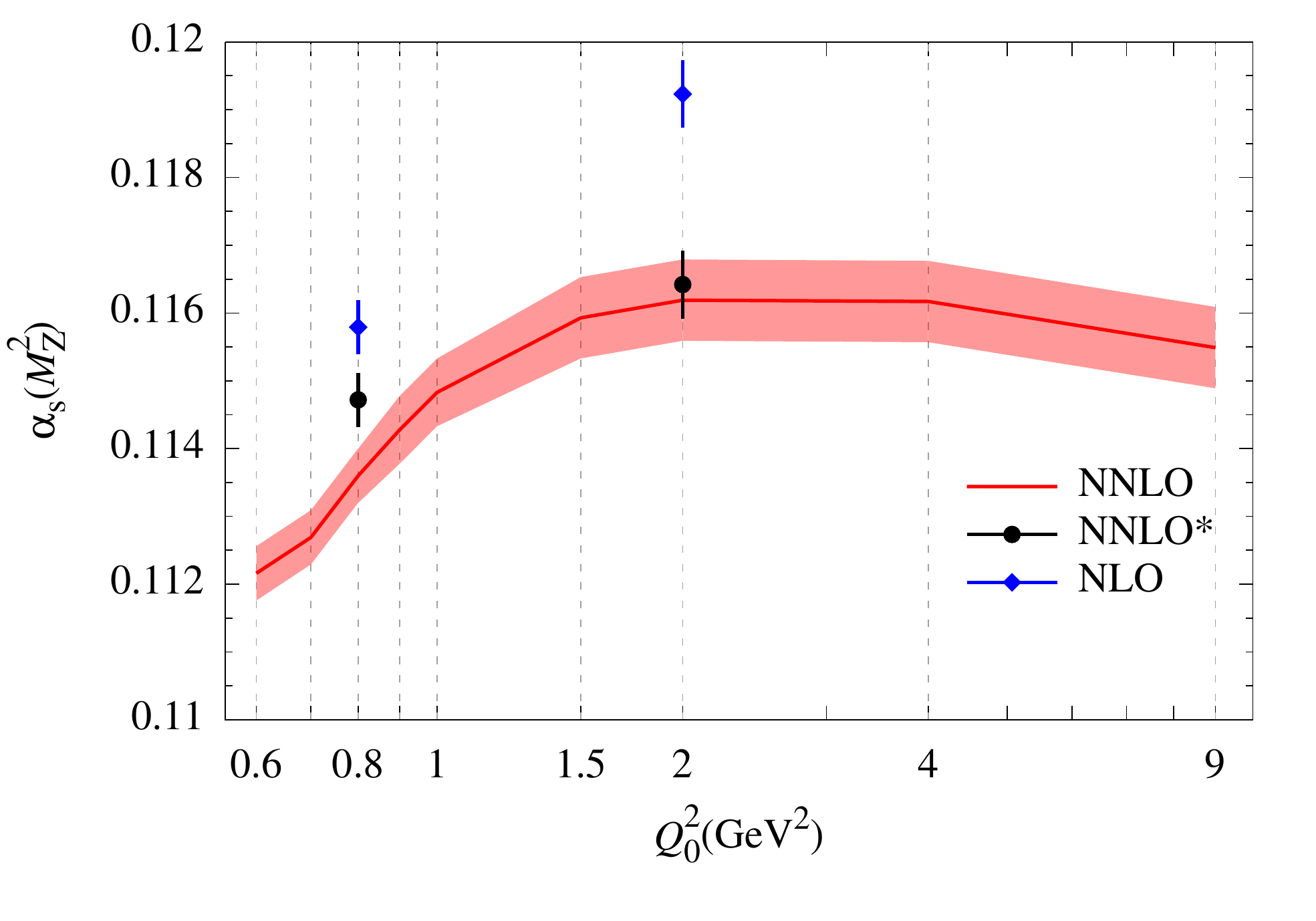}
\caption{The dependence of $\chi^2$ and $\alpha_s(M_Z^2)$ values on the variations of the input scale $Q_0$ as obtained in our NNLO analyses,
together with the $1\sigma$ uncertainty band ($\Delta\chi^2=1$). For illustration we show the sensitivity of two of our results when including jet and
charm data at NNLO as well, denoted by NNLO*, where NLO matrix elements have been (inconsistently) combined with NNLO PDFs. The NLO results at the input scales 0.8 GeV$^2$ (dynamical) and 2 GeV$^2$ (standard) are depicted as well.}
\label{input}
\end{figure*}

An innovation in our global analyses is the explicit study of the dependence of the results on the specific choice of input scale $Q_0^2$, which in most cases had not been systematically addressed so far. As demonstrated in \cite{JimenezDelgado:2012zx}, although in theory the results should not depend on these choices, in practice a relevant dependence develops as a consequence of what has been called \emph{procedural} bias. In principle this uncertainty should be reduced as much as possible, however the remaining procedural uncertainty in the fits can be estimated through variations of the input scale. This point is illustrated in Fig.~\ref{input}
where values of the minimum global $\chi^2$ and the corresponding $\alpha_s$ values are shown. Note that while the fits remain of comparable quality (comparable $\chi^2$) until very low values of the input scale (where perturbation theory starts to break down), the sensitivity of other quantities might be larger, as exemplifies here by $\alpha_s$. Since, except for the $Q_0$ values, the fits are done under the exact same conditions, the inability of the global procedure to distinguish different results is due to shortcomings in the procedure itself, for example difficulties of the parametrizations to produce equally optimal shapes of the distributions at different scale (which are well known to be very different \cite{Gluck:2007ck,JimenezDelgado:2008hf}. To include this uncertainties in our predictions we produce in each case two sets of PDFs, in the so-called dynamical $(Q_0^2 < 1$ GeV$^2$) and standard ($Q_0^2 \gtrsim 1$ GeV$^2$) approaches, thus the uncertainty can be estimated by comparing this two distinct results. In the case of the strong coupling, the results obtained for the central values and the statistical uncertainties (only) from our nominal dynamical NNLO (NLO) analyses are $\alpha_s(M_Z^2) = 0.1136\pm 0.0004$ $(0.1158\pm 0.0004)$, whereas the standard fits give $\alpha_s(M_Z^2)=0.1162\pm 0.0006$ $(0.1191\pm 0.0005)$; so that an additional uncertainty of $\Delta_{\rm proc.}\alpha_s\!=\!0.0013$ can be attributed to procedural uncertainties. Of course, this does not exhaust the list of systematic uncertainties, choices like data selection and genuinely theoretical uncertainties like scheme and scale choices in the analysis should also be considered, and would further increase the total error. 
\begin{figure*}
\centering
\includegraphics[width=0.49\textwidth]{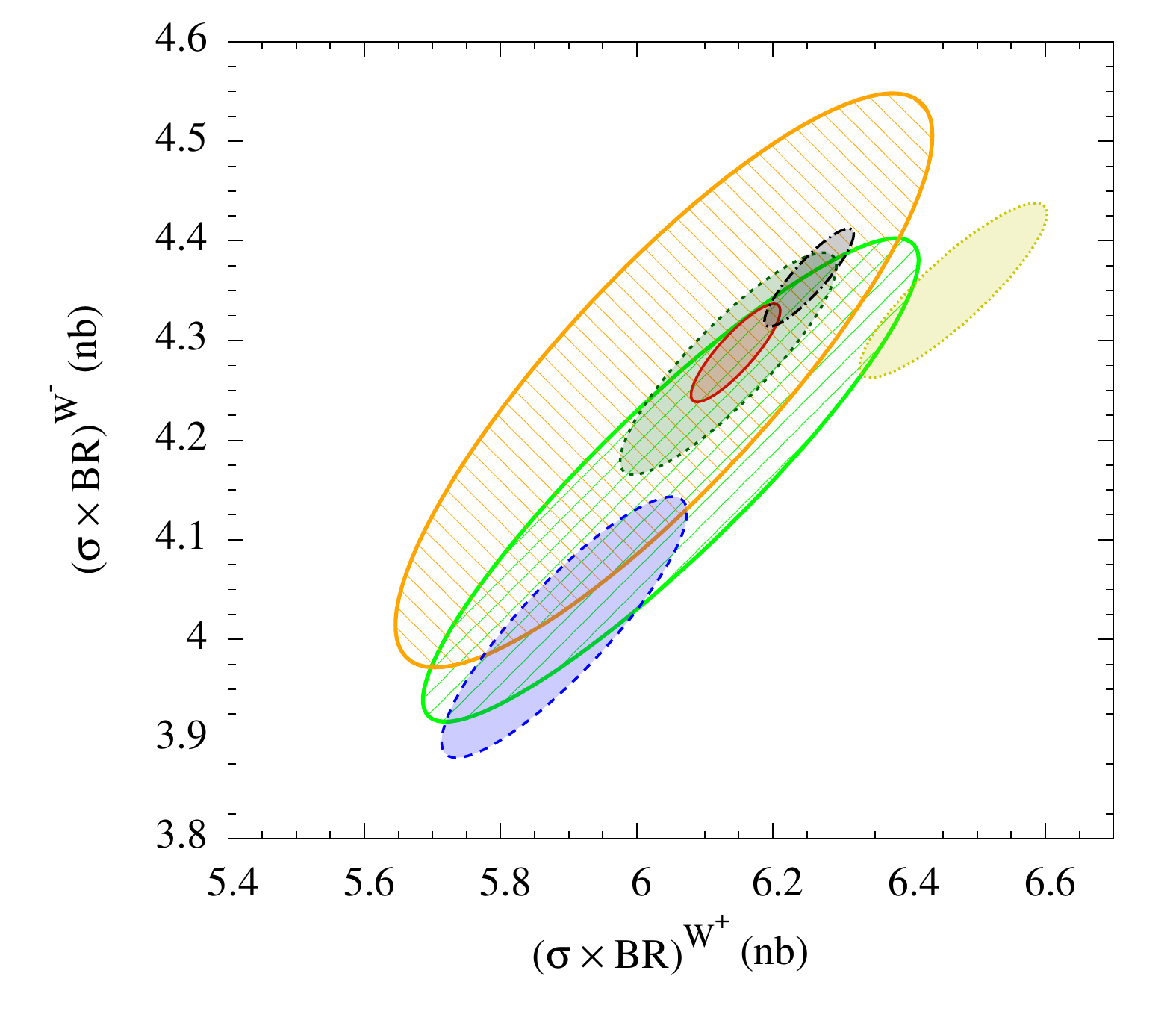}
\includegraphics[width=0.49\textwidth]{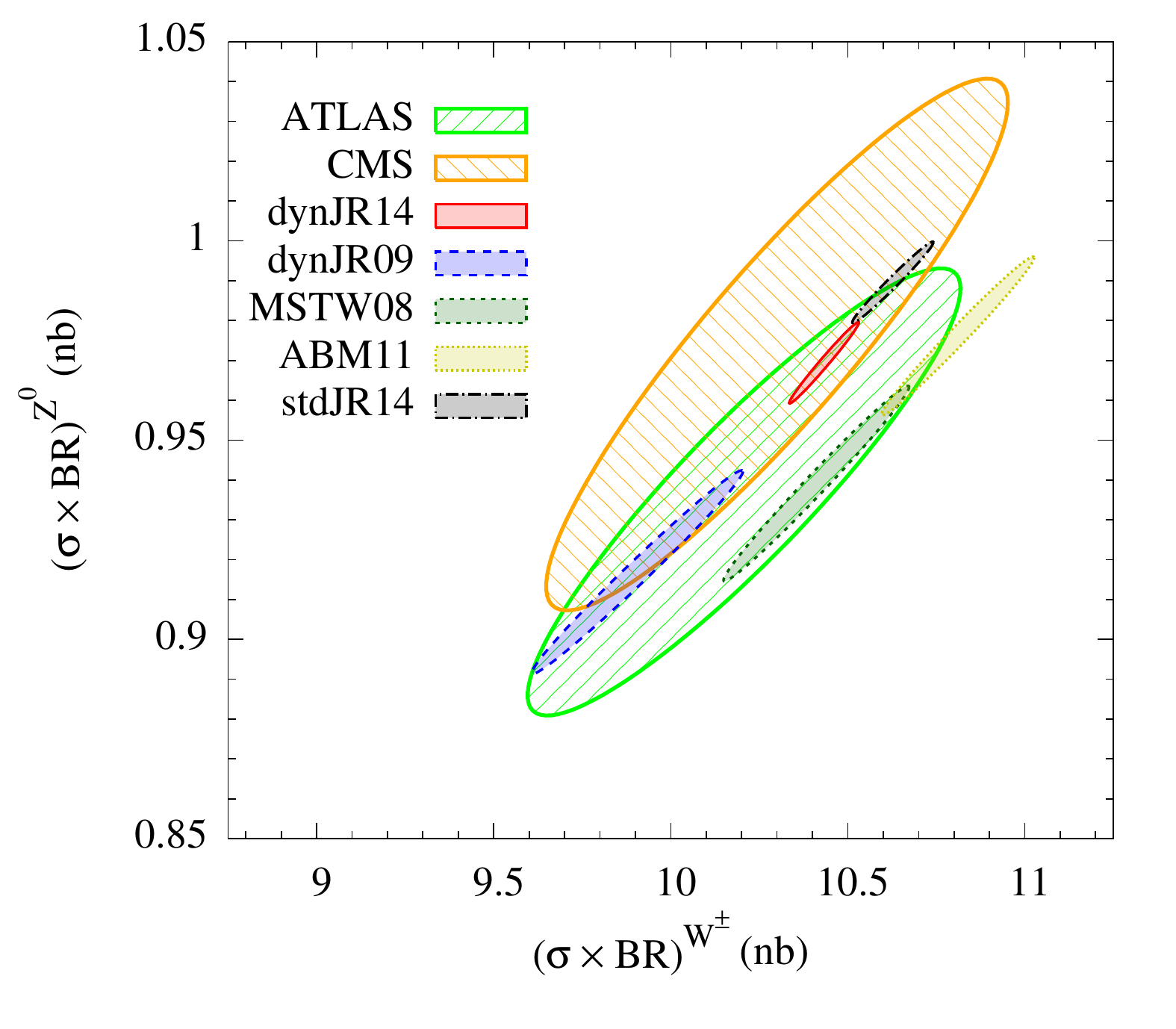}
\caption{1$\sigma$ ($\Delta \chi^2 = 2.3$) ellipses for $W^{\pm}\equiv W^+ +W^-$ vs. $Z^0$ and $W^{+}$ vs. $W^-$ total $pp$ cross sections at NNLO, compared to the LHC data from ATLAS \cite{ref67} and CMS\cite{ref68} at $\sqrt{s}=7$ TeV. For comparison we display the dynamical predictions of JR09 \cite{JimenezDelgado:2008hf} as well as of ABM11 \cite{ref5} and MSTW08 \cite{ref1}.}
\label{benchmarks}
\end{figure*}

Our dynamical and standard results should be relevant for the calculation of cross-sections at ongoing experiments at the LHC. It should be emphasized that, on purpose, we have not included TeVatron gauge-boson production data and LHC data in our fitting procedure, in order to allow for genuine {\em predictions} of these measurements. Detailed benchmark of our results and comparisons with other groups have been presented in \cite{Jimenez-Delgado:2014twa} and cannot be repeated here. However, for illustration, we compare in Fig.~\ref{benchmarks} our dynamical and standard predictions for gauge boson production with LHC data \cite{ref67,ref68} and results from ABM11\cite{ref5} and MSTW08\cite{ref1}. Note that although each of our new results has a relatively small error, due partially to our rather stringent tolerance criteria for 1$\sigma$ parameter errors ($\Delta \chi^2 = 1$), the combination of our dynamical and standard fits provides an estimation of the uncertainties which is closer in side to our previous results (where less stringent tolerance criteria were used), although the improvement in accuracy is also noticeable. 

\section{Polarized case: JAM}\label{polarized}
The first global NLO analysis of spin-dependent PDFs from the JAM collaboration \cite{Jimenez-Delgado:2013boa} utilizes the available data on inclusive polarized DIS from protons, deuterons and $^3$He. Where possible, we have fitted directly the measured polarization asymmetries, rather than relying on structure functions extracted under different conditions than the unpolarized cross sections. We include data from all polarized DIS experiments that lie within the limits $Q^2 \geq 1$~GeV$^2$ and $W^2 \geq 3.5$~GeV$^2$, which allows us to constrain the $\Delta u^+$ and $\Delta d^+$ distributions\footnote{Here and below $q^+ = q + \bar{q}$.} up to $x \approx 0.7$. Obtaining stable fits over this expanded kinematic range necessitates systematically accounting for target mass and higher twist corrections, which are vital for describing the $g_1$ and $g_2$ structure functions at the lower $Q^2$ range, and nuclear smearing corrections for deuterium and $^3$He nuclei, which have major impact at large~$x$.

\begin{figure*}
\centering
\includegraphics[width=0.49\textwidth]{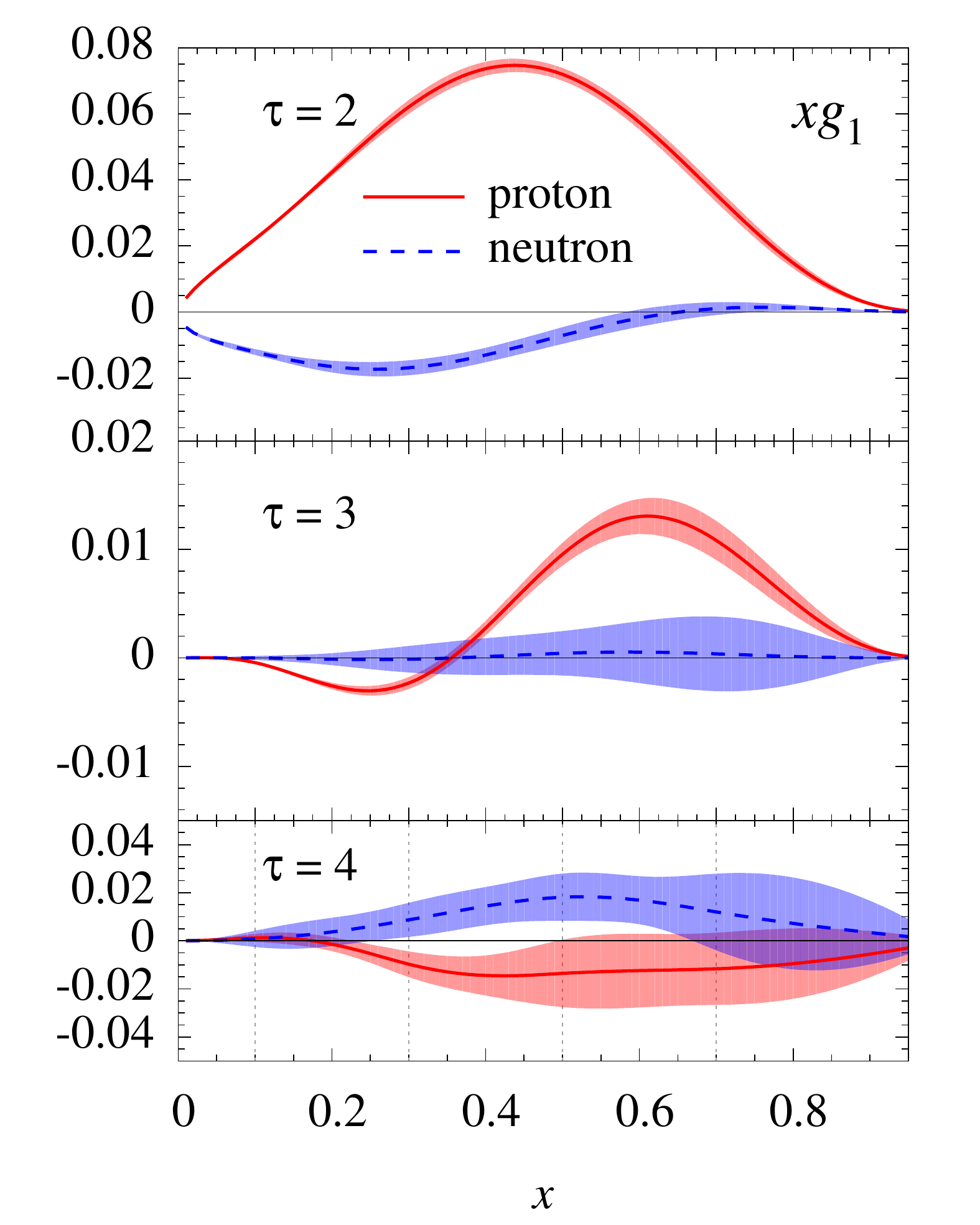}
\includegraphics[width=0.49\textwidth]{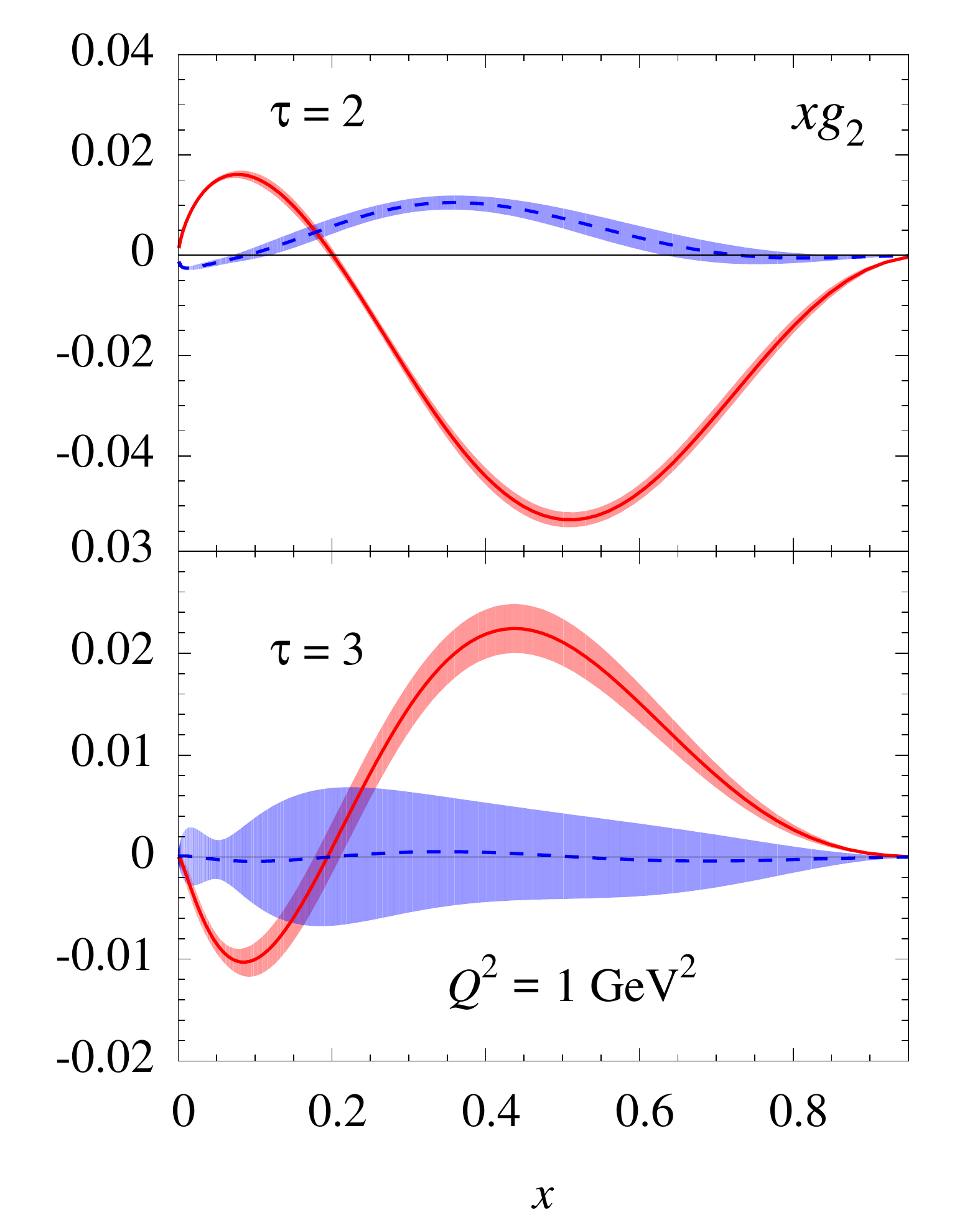}
\caption{Twist decomposition of the proton (red solid) and neutron (blue dashed) $xg_1$ and $xg_2$ structure function. For the $xg_1$ structure function (left) the twist $\tau=2$ (top panel), $\tau=3$ (middle) and $\tau=4$ (bottom) are shown, while for the $xg_2$ structure function (right) the $\tau=2$ (top) and $\tau=3$ (bottom) contributions are illustrated. The dotted vertical lines on the $\tau=4$ contribution to $xg_1$ represent the knots used for the spline fit.}
\label{polarizedHT}
\end{figure*}

The results of the JAM13 fits indicate that the $\Delta d^+$ distribution has a significantly larger magnitude in the intermediate-$x$ region ($x \gtrsim 0.2$) than in previous analyses, due primarily to the sizable higher twist corrections found here and shown in Fig.~\ref{polarizedHT}.
In particular, the twist $\tau=3$ term makes important contributions to both the $g_1$ and $g_2$ structure functions of the proton, and the $\tau=4$ correction makes a large and positive contribution to the neutron $g_1$.  The latter is mostly responsible for driving the $\Delta d^+$ distribution to become more negative. The induced twist-3 contribution to the proton $g_1$ also reduces the size of the twist-4 term compared to that found previously. The $\tau=3$ correction to the neutron $g_2$ is compatible with zero within errors. This clearly highlights the importance of including subleading $1/Q^2$ corrections in any analysis that attempts to fit data down to $Q^2 \sim 1$~GeV$^2$, and even if more stringent cuts are imposed. 

\begin{figure*}
\centering
\includegraphics[width=0.49\textwidth]{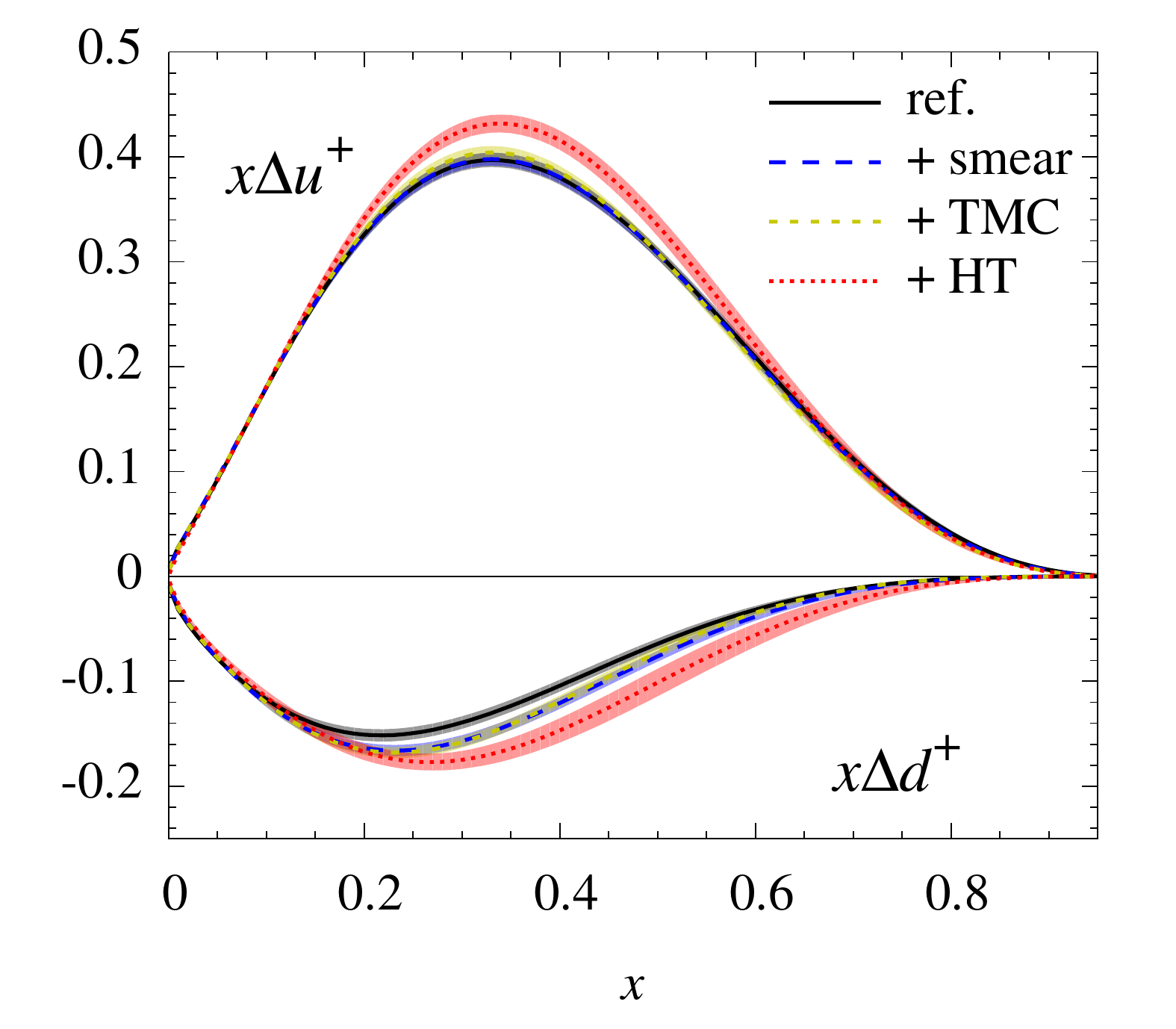}
\includegraphics[width=0.49\textwidth]{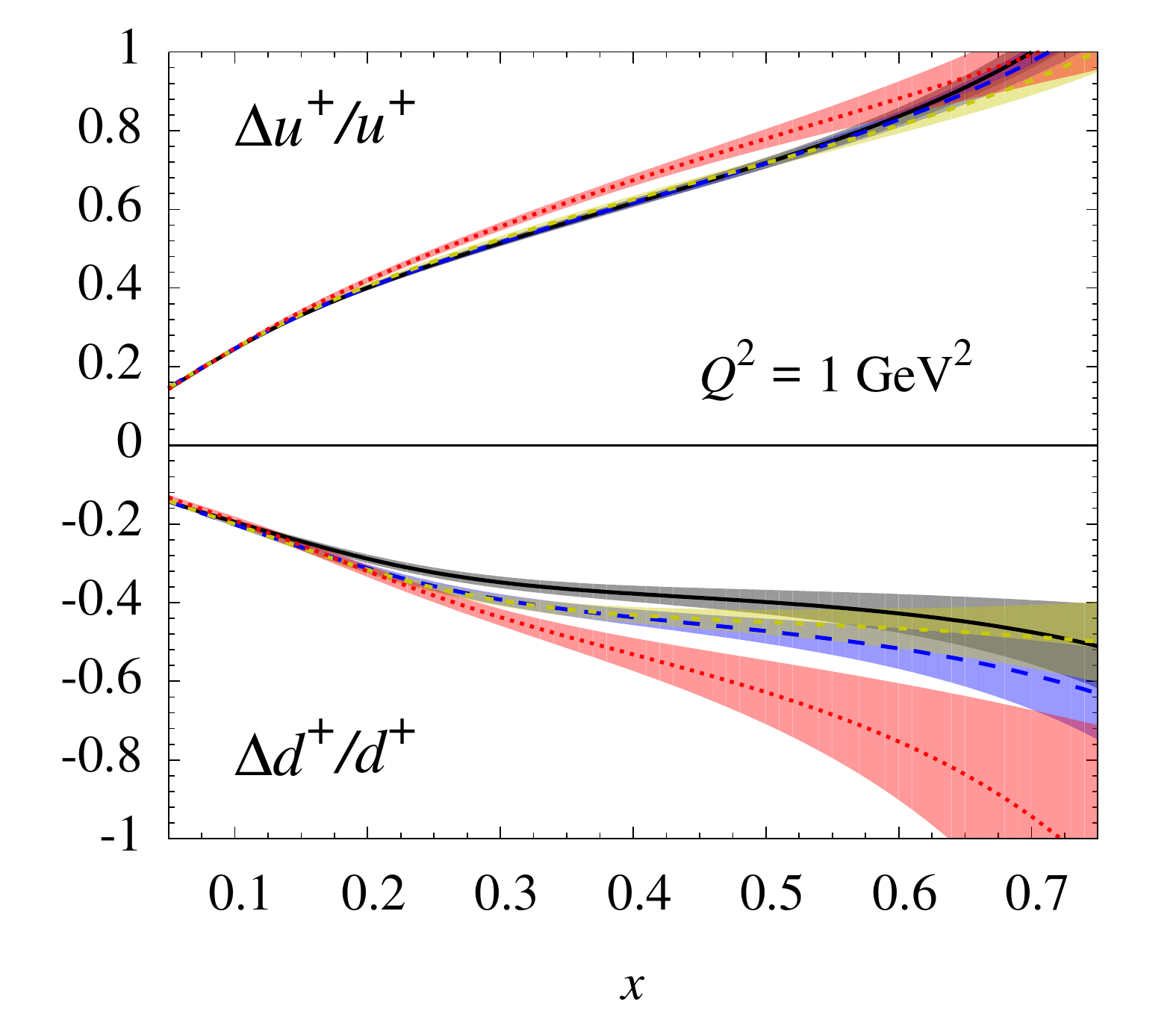}
\caption{(Left) Spin-dependent $\Delta u^+$ and $\Delta d^+$ distributions showing the cumulative effects on the reference PDFs (solid) of adding nuclear smearing (red dashed), target mass (blue short-dashed), and higher twist (green dotted) corrections. (Right) Corresponding ratios of polarized to unpolarized $\Delta u^+/u^+$ and $\Delta d^+/d^+$ distributions.}
\label{polarizedPDFs}
\end{figure*}

An illustration of the effects of the corrections studied in the JAM13 analysis is presented in Fig.~\ref{polarizedPDFs}, where beginning with the reference twist-2 NLO QCD parametrizations (without any additional corrections) the cumulative effects of the nuclear smearing, target mass and higher twist corrections on the $\Delta u^+$ and $\Delta d^+$ distributions are demonstrated explicitly. The impact of these corrections is negligible at small values of $x$, $x \lesssim 0.2$, but grows increasingly important at higher $x$. Compared with the reference distributions, both the JAM $\Delta u^+$ and $\Delta d^+$ PDFs are larger in magnitude, by $\sim 10\%-20\%$ for the $u$ quark at $0.2 \lesssim x \lesssim 0.6$, and by more than $50\%-100\%$ for the $d$ quark at $x \gtrsim 0.4$. The same effects are more clearly illustrated in the form of ratios of polarized to unpolarized PDFs $\Delta u^+/u^+$ and $\Delta d^+/d^+$, shown in Fig.~\ref{polarizedPDFs}(right). Such a comparison is meaningful since the unpolarized PDFs are fitted within the same analysis and applied consistently in the extraction of the polarized PDFs.

We have also investigated whether existing data from polarized lepton--nucleon DIS is able to provide any constraints on the $x \to 1$ behavior of spin-dependent PDFs in the context of a global QCD analysis. Using the JAM13 fit as a baseline, we found \cite{Jimenez-Delgado:2014xza} that demanding the polarized to unpolarized PDF ratios $\Delta q^+/q^+$ to approach unity at $x=1$ results in equally good fits to the available data, even though the resulting changes to the $\Delta d^+$ PDF are significant in the intermediate-$x$ region.  With dramatically different behaviors for the $\Delta d^+/d^+$ ratio allowed for $x \gtrsim 0.3$, this highlights the critical need for precise data sensitive to the $d$ quark polarization at large $x$ values.

Constraining the behavior of the polarization asymmetries $A_1$, and consequently of the spin-dependent PDFs, in the limit as $x \to 1$ is one of the featured goals of the experimental physics program planned for the 12~GeV energy upgraded CEBAF accelerator at Jefferson Lab.  Data from several experiments are expected to be collected for values of $x$ as high as $\approx 0.8$ for DIS kinematics, and even higher $x$ in the nucleon resonance region.  This should significantly reduce the PDF uncertainties for $x \gtrsim 0.5$, especially for the $\Delta d^+$ distribution, which will be more strongly constrained by new data on the $^3$He asymmetry.

\begin{figure*}[b]
\centering
\includegraphics[width=0.49\textwidth]{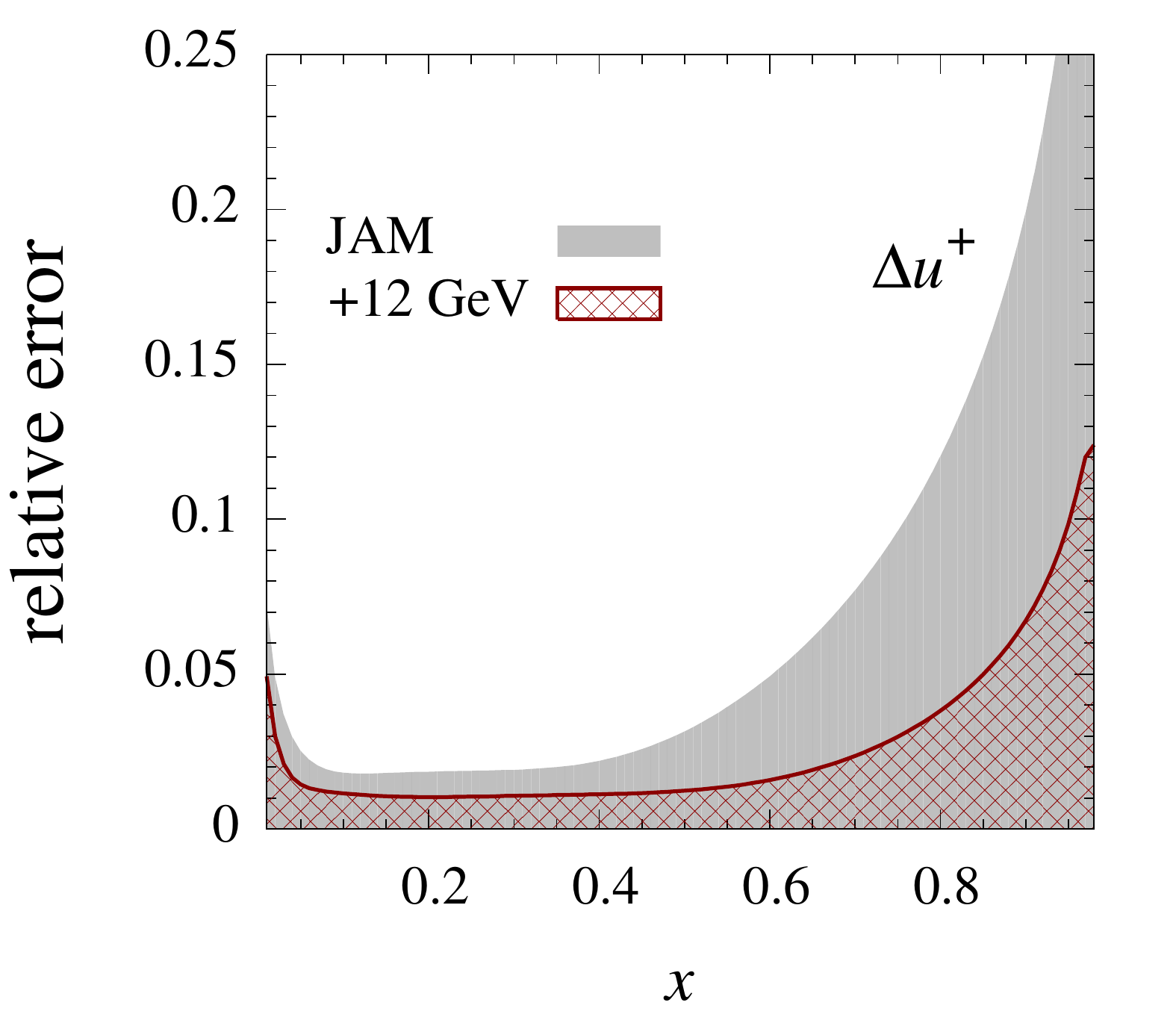}
\includegraphics[width=0.49\textwidth]{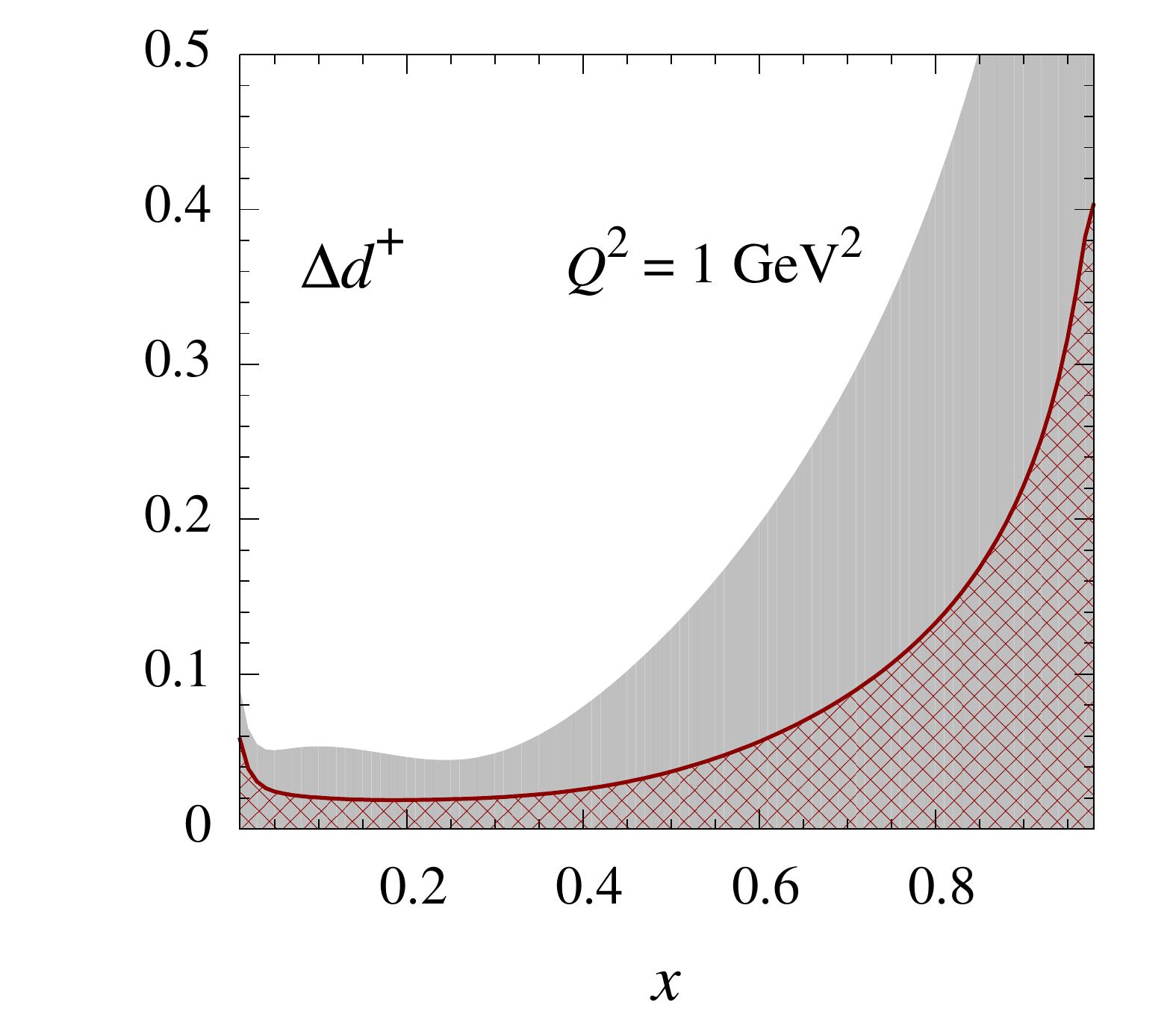}
\caption{Relative error on the $\Delta u^+$ (left) and $\Delta d^+$ (right) PDFs for the JAM fit at $Q^2=1$~GeV$^2$ (gray band) and for JAM including pseudo-data expected from planned Jefferson Lab 12~GeV experiments (red hashed area).}
\label{JLab12}
\end{figure*}
  
To estimate the possible impact of the new Jefferson Lab data we \cite{Jimenez-Delgado:2014xza} use the projected statistical and systematic uncertainties for the proposed experiments at the $x$ and $Q^2$ values where the asymmetries will be measured. The pseudo-data are generated by randomly distributing the central values of the points around the JAM fit for hydrogen, deuterium and $^3$He targets (distributing them around other predictions would be equally suitable). The reduction in the PDF uncertainties, illustrated in Fig.~\ref{JLab12}, is significant, with the relative error on $\Delta u^+$ and $\Delta d^+$ decreasing by $\sim 70\%$ for $x=0.6-0.8$ at the input scale $Q^2=1$~GeV$^2$.

\section*{Summary and Conclusions}\label{conclusions}
We have reported on some aspect of our latest updates of unpolarized \cite{Jimenez-Delgado:2014twa} and polarized \cite{Jimenez-Delgado:2013boa, Jimenez-Delgado:2014xza} PDFs. In both cases we find that the inclusion of data at relatively low $Q^2$ and $W^2$ values, as well as a wealth of data on nuclear targets, provides additional valuable information on the partonic structure of the nucleon. In the polarized case it has helped to increase the statistical accuracy of our previous analyses \cite{JimenezDelgado:2008hf}, while in the polarized case it has lead to important changes in the central values of the distributions. 

In addition it has been shown that rather than arbitrarily increasing the estimated uncertainties by using a tolerance criteria $\Delta \chi^2 > 1$ for the propagation of experimental uncertainties, a viable approach is the estimation of the remaining procedural bias in the global fits through variations of the input scale (dynamical vs. standard approaches). 

To finalize we have discussed the need for precise polarized DIS data at large values of $x \gtrsim 0.3$, especially to determine the $\Delta d^+$ distribution. This is one of the goal of the 12 GeV update of Jefferson Lab and we have shown that the projected experimental uncertainties will be able to reduce current uncertainties by about 70\% in the large-$x$ region.

\begin{acknowledgements}
I thank W.~Melnitchouk and E.~Reya for the fruitful collaborations which have lead to this publication. This work was supported by the DOE Contract No.~DE-AC05-06OR23177, under which Jefferson Science Associates, LLC operates Jefferson Lab.
\end{acknowledgements}

\end{document}